\documentstyle[preprint,aps,prd]{revtex}

\newcommand{\Dph}{\Delta \phi}

\newcommand{\Dps}{\Delta \psi}

\newcommand{\Sq}{S^{(2)}}
\newcommand{\be}{\begin{equation}}
\newcommand{\ee}{\end{equation}}
\newcommand{\bea}{\begin{eqnarray}}
\newcommand{\eea}{\end{eqnarray}}
\newcommand{\al}{\alpha}
\newcommand{\e}{\epsilon}
\newcommand{\nn}{\nonumber}
\begin{document}
\draft

\title{One-Loop Supergravity Corrections to the Black Hole Entropy
and Residual Supersymmetry}

\author{Renata Kallosh,
J. Rahmfeld and
Wing Kai Wong}

\address{Physics Department, Stanford University, \\
Stanford, CA
94305-4060}

\preprint{\noindent
\begin{minipage}[t]{3in}
\begin{flushright}
SU-ITP-97-31 \\
hep-th/9706048
\vspace{1truecm}
\end{flushright}
\end{minipage}}

\maketitle

\begin{abstract} 
We study the one-loop corrections to the effective
on-shell action of N=2 supergravity in the background of the
Reissner-Nordstrom black hole. In the extreme case the contributions
from graviton, gravitino and photon to the one-loop corrections to the
entropy are shown to cancel. This gives the first explicit example of
the supersymmetric non-renormalization theorem for the on-shell action
(entropy) for BPS configurations which admit Killing spinors. We display
the residual supersymmetry of the perturbations of a general
supersymmetric theory in a bosonic BPS background. 

\end{abstract}

\vspace{0.5cm}

It was shown by Gibbons and Hawking \cite{GH} for the Reissner-Nordstrom
black holes and in \cite{KAL92b,KOP} for dilaton black holes that the
Bekenstein-Hawking entropy of black holes can be found by evaluating the
Euclidean on-shell action of the theory in the semiclassical
approximation. The first quantum corrections to the entropy can be found
by evaluating the one-loop partition function in the black hole
background. 

The arguments in favor of a supersymmetric non-renormalization theorem
for the on-shell action (entropy) for configurations which admit Killing
spinors were presented in \cite{KAL92a,KAL92b,Hull}. The proof is based
on Berezin rules of integration in superspace and the fact that
Killing spinors can be related to isometries in fermionic directions in
superspace. The simplest configuration constitutes the extreme
Reissner-Nordstrom (RN) black hole of pure $N=2$ supergravity
\cite{FER76}. The existence of Killing spinors for this solution was
established in \cite{GIB82}. However, there was no explicit one-loop
calculation available to support or disprove the theorem of
non-renormalization of the entropy, not even in the RN case. 

The situation with the one-loop corrections in the $N=2$ theory was
obscured by the existence of the so-called conformal anomalies
\cite{Duff}. The trace anomaly of the one-loop on-shell supergravity in
a gravitational background is given by the following expression:
\begin{equation}
T=  g^{\mu\nu} < T_{\mu\nu}> \  = g^{\mu\nu} {\partial \over \partial
g^{\mu\nu}} \left\{ { (\det  D_{3/2})^2 \over \det  D_{2} \det  D_{1} }
\right\}= \, {A\over 32 \pi^2} {}^*
R_{\mu\nu\lambda\delta} {}^{*}R^{\mu\nu\lambda\delta} \ .
\label{anomaly}
\end{equation}
Here $D_{s}$ denotes the contribution from positive and negative
helicity states of spin $s=2$, two spin $s=3/2$, and spin $s=1$ fields
of the full $N=2$ supergravity multiplet. The coefficient $A$ is known
for all fields interacting with gravity. 

The one-loop counterterm is proportional to the Euler number of the
manifold, $ S_{\rm one-loop} = {1\over \epsilon} A \, \chi .$ The
fields of $N=2$ supergravity include a graviton, 2 gravitini
and a vector field. The anomaly coefficient $A= {11\over 12}$ of pure
$N=2$ supergravity does not vanish. This means that in a {\it purely
gravitational} background the contributions from spin 2, two spin 3/2,
and spin 1 fields of the full $N=2$ supergravity multiplet do not cancel
in the first loop. Although those backgrounds do not admit Killing
spinors, hence are not relevant for the proving or disproving the
non-renormalization theorem, there was not much incentive in calculating
one-loop corrections in pure $N=2$ supergravity in the background of the
extreme Reissner-Nordstrom black hole. On the other hand, in the $N=4$
theory where the conformal anomalies are absent, the calculations may be
much more complicated and they have not been done either. In this paper,
we will see that the one-loop contributions within the supergravity
multiplet cancel in a background admitting a Killing spinor. 

Meanwhile from a completely different perspective, a study of
quasinormal modes of various black holes was developed (see
\cite{CHA83}). Quasinormal modes of black holes provide an opportunity
to identify black holes when large-scale laser-interferometric detectors
for gravitational waves will be available. The formalism was applied in
the study of black holes with general dilaton coupling constant $a$ in
\cite{Wilczek}. In most cases only numerical calculations are possible.
By applying those methods to the RN black hole H. Onozawa, T. Mishima,
T. Okamura, and H. Ishihara \cite{ONO96} discovered a curious fact: the
resonant frequencies of $s=2$ waves with multipole index $j_s$ coincide
with those of $s=1$ waves with multipole index $j_s-1$ in the extremal
limit. This would be very difficult to establish by looking directly
into the standard form of the wave equations in generic, non-extreme RN
as given in the book of Chandrasekhar \cite{CHA83} in ch. 5, eq. (270).
The potential in the wave equation for $s=1,2$ (photons and gravitons)
is defined there in terms of two different functions $f_1$ and $f_2$ and
two different parameters $q_1$ and $q_2$\footnote{The notation for $q_1$
and $q_2$ is reversed in \cite{CHA83}.} 
\begin{equation}
V_s^{\pm}    = \pm q_s {d f_s \over  dr_*} + q^2_s f^2_s +  [A_s
(A_s -2)] f_s
 \qquad s =1,2.
\label{potential}
\end{equation}
Here
\begin{eqnarray}
f_s&=&  {\Delta \over r^3 [ (A_s-2)r + q_s]}
\end{eqnarray}
and $q_s, A_s, \Delta $ are defined below. The graviton and photon
perturbations are encoded in functions $Z_1$ and $Z_2$ \cite {CHA83}
which satisfy the radial equations\footnote{The radial equations here
describe the axial perturbations with $V_s^{-}\equiv V_s $. The polar
perturbations have potentials $V_s^{+}$. We will rely on the fact
established in \cite{CHA83} that the quasinormal modes for axial and
polar perturbations for gravitational waves (and for electromagnetic
ones) are identical and therefore we will study here only the mechanism
of cancellation in the one-loop approximation of axial perturbation from
spin 1, 3/2, 2. However, a more detailed analysis of this point would be
desirable.}
\begin{equation}
 {\cal D}_s \;  Z_s(r) \equiv \left[ \frac{d^2}{dr^2_*}+ \omega^2
-V_s(r)
\right] Z_s(r) =0,
  \label{eq:rw}
\end{equation}
where
\begin{eqnarray}
  \frac{dr}{dr_*} & = & \frac{\Delta}{r^2},   \qquad \Delta  =
r^2-2Mr+Q^2,\\
 \label{pot} V_s    & = & \frac{\Delta}{r^5} \left[ A_s r
                  - q_s +\frac{4Q^2}{r}\right], \\
   q_1    & = & 3 M- \sqrt{9 M^2+4Q^2 (A_2-2)}, \\
   q_2     &=&  3 M+ \sqrt{9 M^2+4Q^2 (A_1-2)},\\
  A_s      & = & j_s(j_s+1), \\
          j_s&=&l+s,\hspace{1cm} (l=0,1,2,....).
\end{eqnarray}
Here $M$ and $Q$ are the mass and charge of the black hole,
and $j_s$ is an angular multipole index of the perturbation.
In the Schwarzschild case $Q=0$,
$s=1$ and $s=2$ correspond to the electromagnetic
and the gravitational perturbation, respectively. For non-vanishing
charge however, $Z_1$ and $Z_2$ are linear combinations of graviphoton
and graviton perturbations of the same multipole index $j_s$.
Schematically,
this can be understood from the quadratic action
\begin{equation}
S^{(2)}={1\over 2} \left (\Delta g \frac{\delta^2 S}{\delta g \delta g} 
\Delta g
+2 \Delta g \frac{\delta^2 S}{\delta g \delta A} \Delta A+
\Delta A \frac{\delta^2 S}{\delta A \delta A} \Delta A+
\Delta \Psi \frac{\delta^2 S}{\delta \Psi \delta \Psi} \Delta \Psi \right),
\end{equation}
where the derivatives are evaluated using the background configuration.
For Schwarzschild black holes $\frac{\delta^2 S}{\delta g \delta A}$
vanishes and no diagonalization of above action is needed. However, when
$Q\neq 0$, the decoupled variables $Z_1$ and $Z_2$ are linear
combinations of $\Delta g$ and $\Delta A$. 

The spin $s=\frac{3}{2}$ contribution comes purely from the gravitino
and obeys eq.(\ref{eq:rw}) with the potential \cite{AIC81,TOR92}
\begin{eqnarray}
  V_{\frac{3}{2}}      & = & G-\frac{dT_1}{dr_*}, \qquad
  G       =  \frac{\Delta}{r^6} ( \lambda r^2 + 2 M r - 2 Q^2), \\
  T_1    & = & \frac{1}{F - 2Q}\left[\frac{dF}{dr_*} - \lambda
  \sqrt{\lambda^2 +1 }
       \right], \\
  F      & = & \frac{r^6}{\Delta^{1/2}}G, \qquad
 \lambda = (j_\frac{3}{2}-\frac{1}{2})(j_\frac{3}{2}+\frac{3}{2}), \\
 &&  j_{\frac{3}{2}}=l+\frac{3}{2}, \hspace{1cm} (l=0,1,2,...).
\end{eqnarray}

The one-loop corrections to the partition function for the
non-extreme RN black
hole can be calculated using these wave equations. The total answer
will be given by the contribution of each field in the multiplet
for each
$l=0,1,2,...$
\begin{equation}
I_{\rm one-loop} = (e^{i W}) _{\rm one-loop}  =
\prod_{l=0}^{\infty}{}\left\{ {
(\det  {\cal D}_{3/2})^2 \over \det  {\cal D}_{2} \det  {\cal
D}_{1} } \right\}.
\label{oneloop}\end{equation}
Here, we have doubled the contribution from the axial perturbations to
include the polar perturbations. Those have the same quasinormal
frequencies and transmission and absorption coefficients (up to a phase)
\cite{CHA83} which leads us to expect that they will give the same
contributions to the one-loop action as the axial modes. 

For black holes which are far from extreme there is no reason to expect
any cancellation between different spin fields. The potentials are
completely different. In particular, if one would take $Q=0$
corresponding to the Schwarzschild black hole it is reasonable to expect
that the conformal anomaly expression in eq. (\ref{anomaly}) will be
reproduced. This is possible since the Euler number of the Schwarzschild
black hole equals 2. It is difficult however to get anything else from
the study of the expression for the one-loop corrections for the
non-extreme black holes. 

Looking into either form of the potential for spin one and spin two
waves as given in equations (\ref{potential}) and (\ref{pot}) for
non-extreme black holes, as well as on the gravitino potential, it is
difficult to see how they could be related in the extreme limit. In this
limit at $M=1$ we get
\begin{equation}
q_1= 4+ 2j_1 \qquad q_2 = 2-2j_2
\end{equation}
But even at extreme $V_1$ is different from $V_2$ for all $l$:
\begin{equation}
V_1-V_2= \frac{2 \Delta}{r^5} (2-r) (l+2).
\end{equation}
However, {\it numerical calculations} in \cite{ONO96} have shown that
the quasinormal frequency trajectories of the photon, gravitino and
graviton with increasing charge meet at the same point in the limit of
maximal charge $Q=M$. Moreover, in \cite{AO} later the numerical
calculations were extended to higher {\it rapidly damped modes} of
nearly extreme black holes and it was found that they {\it spiral into
the value of the extreme black hole as the charge increases}. The
behavior of the trajectories was found to change dramatically for the
charge $M\geq Q> 0.9 M$. 

To explain this curious behavior of quasinormal modes for the various
spin waves in the background of the near extreme RN black holes, the
authors of \cite{ONO96} and \cite{ONO97} were able to reorganize the
form of the potentials in the wave equations in such a way as to explain
their previous results in an analytic way for the extreme black holes
with $M=Q=1$ with horizon at $r=1$. First of all the potential for all 3
fields is now given in terms of only one function of $r$
\begin{equation}
  f = \frac{r-1}{r^2}.
\end{equation}
The potentials for the extreme RN solution acquire the form
\begin{eqnarray}
    V_1 &=& + (j_1+1) \frac{df}{dr_*}
   -4  f^3 + (j_1+1)^2 f^2,     \label{eq:v1}  \\
   V_{\frac{3}{2}} &=& + (j_\frac{3}{2} + \frac{1}{2})\frac{df}{dr_*}
                          -4  f^3 + (j_\frac{3}{2}+\frac{1}{2})^2 f^2,
   \label{eq:v32} \\
   V_2 &=& - j_2 \frac{df}{dr_*}
         -4  f^3 + j_2^2 f^2,
   \label{eq:v2}  \\
&&  j_s=l+s, \hspace{1cm} (l=0,1,2,...),
\end{eqnarray}
and the difference between $V_1$ and $V_2$ becomes
\begin{equation}
V_1-V_2= \frac{2 \Delta}{r^2} (2-r) (l+2)= (j_1+j_2+1)
\frac{df}{dr_*} \ .
\end{equation}
The tortoise coordinate  $ r_*$ in terms of which the wave differential
operator is very simple for the extreme RN black hole is
\begin{equation}
  r_* = r-1 + \ln (r-1)^2 - \frac{1}{r-1}.
\end{equation}
Of crucial importance is that $r_*$ maps infinity to infinity
and the horizon to negative infinity:
\begin{eqnarray}
r \longrightarrow \infty &\hskip 0.5 cm  \Longrightarrow \hskip
0.5 cm &r_* \longrightarrow \infty \\
\nonumber\\
r \longrightarrow 1 &\hskip 0.5 cm  \Longrightarrow \hskip 0.5 cm
& r_* \longrightarrow  -\infty.
\end{eqnarray}
Also the relation
\begin{equation}
r_*\longrightarrow -r_* \hskip 0.5 cm  \Longrightarrow \hskip 0.5 cm
r-1 \longrightarrow \frac{1}{r-1} \label{parity}
\end{equation}
will prove to be very useful.

Fortunately for supersymmetry, as we will see later, the wave equation
for the radial coordinate has a simple second order differential
operator in tortoise coordinate as given in $ {\cal D}_s$ in eq.
(\ref{eq:rw}). 

These three potentials are related in the following way:
\begin{equation}
  V_1(r_*,j_1=j)=V_{\frac{3}{2}}(r_*,j_{\frac{3}{2}}=j+\frac{1}{2})=
                V_2(-r_*,j_2=j+1),
      \label{eq:mr}
\end{equation}
where $j$ is a positive integer. The first equality is obvious in eqs.
(\ref{eq:v1}) and (\ref{eq:v32}); the potential of the Rarita-Schwinger
field is identical to that of the spin one field if we shift the
multipole index by 1/2. The second equality is proved in \cite{ONO97} by
using the relation $f(r_*)=f(-r_*)$ which follows from (\ref{parity}).
Therefore $V_2$ can be obtained by reflecting $V_1$ or $V_{\frac{3}{2}}$
about $r_*=0$. {\it This transformation corresponds to the exchange of
the horizon and infinity}. It has been deduced from eq. (\ref{eq:mr}) in
\cite{ONO97} that a scattering problem for each perturbed field with a
corresponding multipole index results in the same transmission and
reflection amplitudes. This was also derived very recently using the
unbroken supersymmetry of the extreme RN in \cite{Okamura}. 

Where does all this bring us with respect to the one-loop corrections to
the entropy of the extreme black holes? If not for the strange fact that
the potential for the $s=2$ perturbations are related to the $s=1$
perturbations by the inversion of the tortoise coordinate, we would not
be able to conclude that there are no corrections. However, the fact
that in calculating one-loop Feynman diagrams in the background of the
extreme black hole an exchange of the horizon with infinity is
needed is rather unusual. Let us look into this closely. We would like
to compare the one-loop contribution for every $l$ for $s=1$ and $s=2$
perturbations. The relevant part of the path integral is
\begin{equation}
\int d \Phi_1 d \Phi_2 \exp {i \int dr_* (  \Phi_1 {\cal D}_{1}
\Phi_1  +
\Phi_2  {\cal D}_{2} \Phi_2)}
\end{equation}
where
\begin{equation}
{\cal D}_{1}=  \left[ \frac{d^2}{dr^2_*}+ \omega^2 -V_1(r_*) \right],
\hskip 1 cm
{\cal D}_{2}=  \left[ \frac{d^2}{dr^2_*}+ \omega^2 -V_2(r_*) \right].
\end{equation}
Now we may use the facts that the term $\frac{d^2}{dr^2_*}$ is even
in $r_*$
and that $V_1(r_*)=V_2(- r_*)$ for the same value of $l$.
Hence, we conclude
that
\begin{equation}
 {\cal D}_{1} (r_*) =    {\cal D}_{2} (- r_*),
\end{equation}
i.e. not only the potential but also the total wave 
equations for spin one and spin two are related by the inversion 
of the tortoise coordinate and
therefore by the exchange of the horizon and infinity. Thus the
quadratic part of the relevant action for $s=1$ is
\begin{equation}
S_1 = \int _{-\infty} ^{+\infty} dr_*   \Phi_1(r_*)  {\cal D}_{1} \left
(
\frac{d}{dr_*} , r_* \right)   \Phi_1(r_*)
\end{equation}
and the quadratic part of the relevant action for $s=2$ is
\begin{equation}
S_2 = \int _{-\infty} ^{+\infty} dr_*   \Phi_2 (r_*)  {\cal D}_{2}
\left (
\frac{d}{dr_*} , r_* \right)   \Phi_2(r_*) = \int _{-\infty}
^{+\infty} dr_*
\Phi_2 (r_*)  {\cal D}_{1} \left ( - \frac{d}{dr_*} , - r_* \right)
\Phi_2(r_*).
\end{equation}
Since the integral over $r_*$ extends from $-\infty$ to $+\infty$
we may change
the integration $ r_* \rightarrow  -r_*$  variables and get
\begin{equation}
S_2 =  \int _{-\infty} ^{+\infty} dr_*   \Phi_2 (- r_*)  {\cal
D}_{1} \left (
\frac{d}{dr_*} ,  r_* \right)   \Phi_2(- r_*).
\end{equation}

The last step is to consider the change of variables in the path
integral
and to integrate instead of $ \Phi_2 (r_*) $ over $ \Phi_2 (-r_*)  $
\begin{equation}
\int d \Phi_1(r_*)  d \Phi_2 (-r_*)   \exp {i \int dr_* (
\Phi_1 (r_*) {\cal D}_{1}  \Phi_1 (r_*)  + \Phi_2(-r_*)  {\cal
D}_{1} \Phi_2
(-r_*) )}.
\end{equation}
Introducing the notation
\begin{equation}
\Phi_2(-r_*) = \tilde \Phi_2(r_*)
\end{equation}
we may rewrite the path integral in the form where the contribution
from the
gravitons and from the photons is the same for each $l$.
\begin{equation}
\int d \Phi_1(r_*)  d \tilde \Phi_2 (r_*)   \exp  {i
\int dr_* (
\Phi_1 (r_*) {\cal D}_{1}  \Phi_1 (r_*)  + \tilde \Phi_2(r_*)
{\cal D}_{1}
\tilde \Phi_2 (r_*) )}
\end{equation}
Since the wave equation for the radial mode of the gravitino for each
$l$
coincides with the equation for $s=1$ we conclude that we have
now proved
the total (axial)
cancellation of the one-loop correction to the on-shell
action of $N=2$ supergravity in the extreme RN background. Assuming
that the polar modes give the same contributions as the axial ones
we find
\begin{equation}
I_{\rm one-loop} = (e^{i W}) _{\rm one-loop}  =
\prod_{l=0}^{\infty}{}\left\{ {
(\det  {\cal D}_{3/2})^2 \over \det  {\cal D}_{2} \det  {\cal D}_{1} }
\right\}=
\prod_{l=0}^{\infty}{}\left\{ { \det  {\cal D}_{3/2} \over \det
{\cal D}_{1}}
\right\}^2=1.
\end{equation}
The non-trivial part of this analysis was the use of the coordinate
system where the horizon of the extreme black hole is pushed to infinity
which allows a particular change of variables in the one-loop path
integral. 

Thus, the cancellation of the one-loop corrections to the on-shell
action (black hole entropy) found here supports the general argument of
the supersymmetric non-renormalization theorem
\cite{KAL92a,KAL92b,Hull}. The theorem states that the extreme black
hole entropy does not have corrections in any loop order in supergravity
theory. This is in complete agreement with the topological character of
the entropy of black holes with unbroken supersymmetry \cite{FKS}.
\bigskip 

It would be interesting to see whether a similar cancellation can be
established in extremal but non-supersymmetric black holes as suggested
in \cite{Duff}. However, the perturbation equations will be very
different and without the hidden help of unbroken supersymmetries it is
not clear whether the arguments presented in this paper can be repeated.

\bigskip 

The striking similarity between the spin $1,\frac{3}{2}$ and spin $2$
perturbations in the extremal supersymmetric case certainly has a deeper
explanation in the unbroken supersymmetry of the background. In the
present case this was established in \cite{Okamura}. 

We will now set up a {\it general framework to relate various
perturbations via the unbroken residual supersymmetry}. As the Killing
spinor, which is the transformation parameter, is neither constant nor
an arbitrary function in supergravity, we are dealing with {\it rigid
supersymmetry}. We are using here the condensed notation introduced by
B. De Witt \cite{dewitt} and developed in the context of the background
field method for gauge theories in \cite{K74}. The underlying action
$S(\phi,\psi)$ depending on bosonic and fermionic fields $\phi^i$ and 
$\psi^a$
is invariant under the supersymmetry transformations
\bea
\delta \phi^i & = & R^i_\al(\phi,\psi) \e^\al \nn \\
\delta \psi^a & = & R^a_\al(\phi,\psi) \e^\al \label{fullsusy}
\eea
where $R$ denote field dependent matrices and $\e^\al$ are the
supersymmetry parameters. If we are interested in bosonic and fermionic
perturbations $\Dph^i$
and $\Dps^a$ propagating in a background given by $\phi_0,\psi_0$
it is useful to expand the action and the supersymmetry transformations.

For a purely bosonic background with
$\psi_0=0$ the expansion of the action is given
by
\be
S(\phi,\psi)=S^{(0)}(\phi_0,\psi_0=0)
+S^{(2)}(\Dph,\Dps,\phi_0,\psi_0=0)+...
\ee
where $S^{(2)}$ is the Gaussian action for the perturbations given by
\be
S^{(2)}=
\frac{1}{2} \Dph^i S_{,ij}^{(0)} \Dph^j
+\frac{1}{2}\Dps^a S_{,ab}^{(0)} \Dps^b .
\ee
This action eventually gives rise to eqn. (4) and also governs the
one-loop corrections. The indices denote derivatives with respect to
bosons ($i,j...$) and fermions ($a,b,...$). The superscript $\phantom{
}^{(0)}$ implies that the relevant expressions are evaluated using the
background fields and summation is understood to include integration. 

The supersymmetry transformations become
\bea
\delta (\phi^i+\Dph^i) & = & R^{(0) i}_\al \e^\al +
         R^{(0) i}_{\al,a}\Dps^a \e^\al \nn \\
\delta (\psi^a+\Dps^a) & = & R^{(0)  a}_\al \e^\al +
         R^{(0)a}_{\al,i}\Dph^i \e^\al .
\eea
$S(\phi,\psi)$ is clearly invariant under the full set of
transformations,
however we are interested in the symmetries of $\Sq$ in a fixed
background. This restricts the transformations to those which leave
the background invariant. The remaining symmetry of $\Sq$ is a rigid
supersymmetry where the transformation parameters are the Killing
spinors $\e_K^A$ of the background. The resulting symmetry
transformations on the perturbations are
\bea
\delta \Dph^i & = &  R^{(0)  i}_{A,a}\Dps^a \e_K^A \nn \\
\delta \Dps^a & = &  R^{(0)  a}_{A,i}\Dph^i \e_K^A
\label{residual}\eea
where the index $A$ denotes the Killing spinors, rather than the
whole set of transformation parameters.
These equations relate bosonic and fermionic perturbations or
quantum fluctuations. Hence, this residual supersymmetry controls
quantum corrections around black holes which admit Killing spinors.

Note that the same formalism applies (in any dimension) to local as well as to
global supersymmetries. Then, in the
local case the  Killing spinors and the
residual supersymmetry are rigid, whereas in
the global case they are
global.


We do not attempt here to make the connection between this general
formalism and eqs. (4)-(10) more precise. The advantage of using eqs.
(4)-(10) is that they are very simple. The advantage of using eqs.
(\ref{residual}) is that they are universal and always {\it connect
solutions for bosonic perturbations with those of fermionic
perturbations via Killing spinors}. This explains the cancellation of
the one-loop corrections to the entropy via residual supersymmetry. 

\bigskip 

\section*{Acknowledgments} We are grateful to H. Onozawa and N.
Andersson for useful correspondence, and the organizers and
participants of the ``Black Hole" conference in Banff and especially G.
Gibbons, G. Horowitz and E. Poisson for stimulating discussions. The
work of R.K., J.R. and W.K.W. is supported by the NSF grant PHY-9219345.

\end{document}